\documentclass[aps,prd,twocolumn,amsmath,amssymb,amsfonts,nofootinbib,superscriptaddress,altaffilletter]{revtex4-1}
\usepackage{array}
\newcolumntype{P}[1]{>{\centCas Aering\arraybackslash}p{#1}}
\usepackage{graphicx}
\usepackage{cellspace}
\usepackage{amssymb}
\usepackage{amsmath}
\usepackage{longtable}
\usepackage{tabu}
\usepackage{ulem}
\usepackage{xcolor}
\usepackage{etoolbox}
\usepackage{supertabular}
\usepackage{dcolumn}
\usepackage{ wasysym }

\def\EatHs{Einstein@Home }

\newcommand{\avgSeg}[1]{\overline{#1}}			

\newcommand{\Freq}{f}
\newcommand{\fdot}{{\dot{\Freq}}}

\newcommand{\Gauss}{\mathrm{\MakeUppercase{G}}}
\newcommand{\Signal}{{\mathrm{\MakeUppercase{S}}}}
\newcommand{\Line}{{\mathrm{\MakeUppercase{L}}}}
\newcommand{\Transient}{{\mathrm{t\MakeUppercase{L}}}}

\newcommand{\NoisetL}{{\Gauss\Line\Transient}}

\newcommand{\BSNtsc}{{\hat\beta}_{{\Signal/\NoisetL}}}	

\newcommand{\F}{\mathcal{F}}		

\newcommand{\avF}{\avgSeg{\F}}

\newtoggle{fullauthorlist}
\toggletrue{fullauthorlist}

\newtoggle{endauthorlist}
\toggletrue{endauthorlist}
\usepackage[utf8]{inputenc}
\DeclareUnicodeCharacter{2009}{-}
\DeclareUnicodeCharacter{2212}{-}

\begin{document}

\title{ 
Deep learning for clustering of continuous gravitational wave candidates
}


\author{B. Beheshtipour}
\email{b.beheshtipour@aei.mpg.de}
\affiliation{Max Planck Institute for Gravitational Physics (Albert Einstein Institute), Callinstrasse 38, 30167 Hannover, Germany}
\affiliation{Leibniz Universit\"at Hannover, D-30167 Hannover, Germany}

\author{M.A. Papa}
\email{maria.alessandra.papa@aei.mpg.de}
\affiliation{Max Planck Institute for Gravitational Physics (Albert Einstein Institute), Callinstrasse 38, 30167 Hannover, Germany}
\affiliation{Leibniz Universit\"at Hannover, D-30167 Hannover, Germany}
\affiliation{University of Wisconsin Milwaukee, 3135 N Maryland Ave, Milwaukee, WI 53211, USA}

\begin{abstract}

In searching for continuous gravitational waves over very many ($\approx 10^{17}$) templates , clustering is a powerful tool which increases the  search sensitivity by identifying and bundling together candidates that are due to the same root cause. We implement a deep learning network that identifies  clusters of signal candidates in the output of continuous gravitational wave searches and assess its performance. For loud signals our network achieves a detection efficiency higher than 97\% with a very low false alarm rate, and maintains a reasonable detection efficiency for signals with lower amplitudes, i.e. at $\lesssim$ current upper limit values. 

\end{abstract}
\maketitle

\section{Introduction}
\label{intro}

Gravitational waves from binary mergers have ushered us in the era of gravitational wave astronomy \cite{abb16,nit19,lig18}. More signals are expected than the ones identified so far, namely unmodeled gravitational wave bursts from catastrophic events \cite{Abbott:2019prv}, stochastic backgrounds \cite{LIGOScientific:2019vic} and continuous gravitational waves \cite{Lasky:2015uia,Authors:2019ztc,Pisarski:2019vxw,Dergachev:2019oyu,Dergachev:2019wqa}. This paper is about an important step in the analysis of results from broad surveys for continuous gravitational waves. 

Continuous gravitational waves are expected from a rotating compact object when its shape deviates from perfect axisymmetry (``mountains'') or due to a non-axisymmetric motion (as an Ekman flow), to the excitation of long-lived r-modes \cite{Haskell:2015iia}, as well as more exotic scenarios \cite{Pierce:2018xmy, Guo:2019ker, Horowitz:2019pru, Palomba:2019vxe}. In spite of the diversity of the emission scenarios the signals can all be described as ever-lasting nearly monochromatic waves.

Continuous wave searches can be relatively straightforward or rather complex. In the first category we find searches for emission from pulsars whose spin evolution is known from electromagnetic observations. In this case the expected gravitational waveform can be predicted based on the observed spin frequency, and standard techniques can be used \cite{Authors:2019ztc,Nieder:2019cyc,Abbott:2019bed}. The computational cost of these searches is negligible and optimal sensitivity can be achieved. The landscape is completely different when one searches for continuous gravitational waves from unseen objects. 

The rationale for searches for continuous gravitational wave signals from unseen objects is solid: whereas we see a few thousand pulsars, based on stellar evolution models, we expect there to be over $10^8$ compact objects in our Galaxy, 10\% of which might be spinning between 10 Hz and 1000 Hz, making them candidates for emission in the high-sensitivity band of LIGO/Virgo. 

The challenge with this type of search is that the number of combinations of frequency, frequency-derivatives and sky positions that one would need to search for, with a fully coherent search over months of data (this would be the optimal strategy), is too high to be doable. So one has to resort to semi-coherent schemes that attain the highest possible sensitivity within the computing budget constraints. There exist a number of semi-coherent approaches, making  different trade-offs between sensitivity, breadth, computational efficiency  and robustness with respect to deviations of the signal waveform from the assumed shape \cite{Walsh:2016hyc}. 

Whatever the specific search may be, the detection statistic may be triggered to rise significantly above the noise by both a gravitational wave or a disturbance. 
Since the template grids are rather fine, whenever the data resembles a template, it is likely that it also resembles nearby templates. This means that for every signal and disturbance we typically have a plethora of templates with detection statistic values above threshold -- ``candidates'' -- to consider. Here ``to consider" means to follow-up with a more sensitive search, to either confirm as interesting or to discard. 

The number of candidates above a given threshold increases as the threshold decreases. The search sensitivity increases as the detection threshold decreases but the total number of candidates that we can follow-up is limited by available computing power. So on the one side we'd like to decrease the threshold to achieve higher sensitivities, on the other we'd like to increase it to keep the computing cost within bounds. 

Clustering helps tip the balance towards higher sensitivities at the same computing cost, by identifying nearby-candidates due to the same root-cause and by bundling them together as one. The follow-up is then performed only based on the most significant of the set, and this reduces the computational cost. 

The clustering algorithms utilised in the highest template-count searches -- the Einstein@Home searches -- have evolved over time \cite{ben15,pap16,sin17}. The latest 
ones measure local over-densities in the detection statistic distribution in parameter space and categorize the resulting clusters as generated by noise or signal, based on their morphology. The morphologies of signals and noise disturbances vary depending on the data and the specific search set-up, so this method requires lengthy preparatory studies and ad-hoc tunings for every search. 

Because so much of the tuning is based on a sort of ``craftmanship'' in interpreting the output of key clustering quantities on fake-signal results and on noise, and based on this, in deducing effective tuning-parameter values, 
we decide to investigate whether a deep learning network could be trained to acquire such craftsmanship and do this just as well or better. In this paper we examine the performance of a clustering method based on a deep learning network.

Deep learning  or deep neural networks are a subfield of machine learning which is inspired by the way brain works.  A neural network consists of a series of neurones (connected processors) that endeavours to recognize relations between a question (input) and an answer (output). This field has greatly developed in the past decade and has proven successful in applications such as image classification, object detection \cite{he17,ron15}, speech recognition and text to speech \cite{hua16,xio16}. 

In physics, and specifically in the field of gravitational waves,  convolutional neural networks 
have been used for data preparation, for instance to classify noise transient in the data \cite{raz18,geo17}  and for de-noising the data of non-gaussian and non-stationary noise \cite{she17, we19}. Neural networks are also being pursued to detect binary inspiral signals and to estimate the associated system parameters \cite{fan19,she19,geb19,geo18,gab18} and recent work has investigated their use for the detection of the elusive continuous gravitational waves \cite{dre19}. 

Armed with this background, we investigate how well a neural network could identify parameter space regions (clusters) in the results of very broad continuous gravitational wave searches, which will be further searched.  We associate the input result data to an image and address the clustering method by using instance segmentation models. 

The paper is organized as follows. Sec.~\ref{meth} describes the network input data and the network architecture. We report on the result of the network in Sec.~\ref{res} and discuss them in Sec.~\ref{Section:summary}.

\section{A neural network for identifying and clustering continuous gravitational wave candidates}
\label{meth}

\subsection{Basics of neural networks}
\label{sec:NNbasics}

Artificial neural networks are computing systems which can be trained to performs certain tasks by example rather than by following an algorithm. The field has a long history \cite{kleene1956} but with increased accessibility to substantial amounts of computing power the last decade has seen renewed interest in artificial neural networks, the blooming of numerous techniques and a slew of applications. 

In this section we introduce some of the key quantities that describe a network and provide an {\it intuitive} description of their role. We have no pretence of providing a thorough review of neural networks and refer the reader to the many existing excellent text books for this (see for example \cite{Goodfellow2015}); this section is for the reader who does not know much about neural networks but would like to understand how we have used them for the problem at hand.  

\begin{figure}[!htb]
\center{\includegraphics[width=0.7\columnwidth]{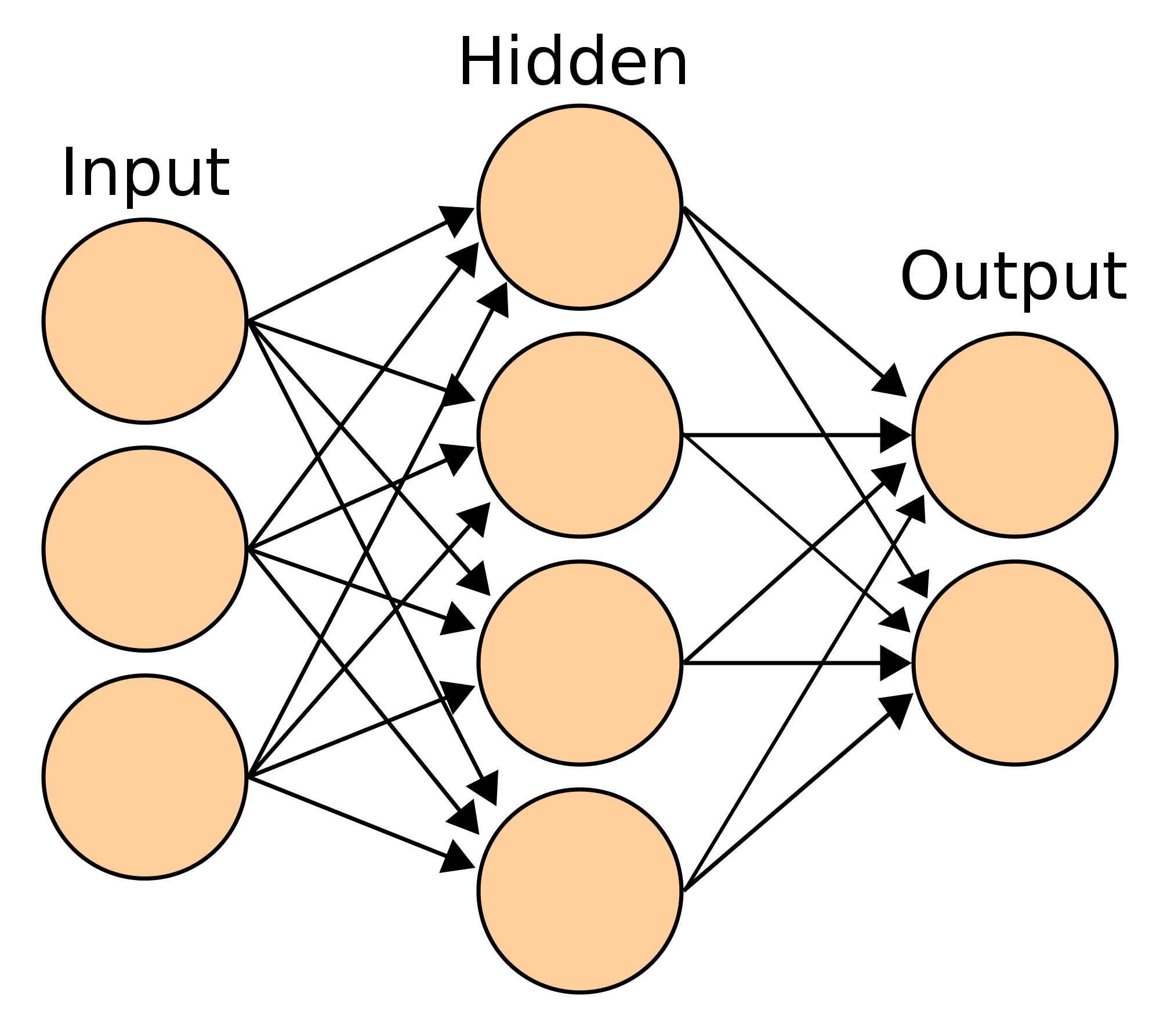}}
\caption{\label{fig:NN} Simplest neural network, consisting of a single hidden layer, from \cite{NNfig}. The nodes (neurones) are represented by the circles. }
\end{figure}

A neural network recognizes relationships/features in the input data and can be used to tell whether some data presents such relationships/features. The network is trained to recognize the desired features based on input data that presents such features. Broadly put, a network labels the data based on various examples of correct labelling of similar data, which are referred to as ``ground truth". 

A network is composed of layers of nodes/neurones. Each node combines the data linearly with a set of coefficients and biases, the so-called model-weights. 

The result of this combination is fully or partially transmitted to the next layer based on its value, through an activation function.

The training consists in determining the sets of coefficients and biases for every layer that result in the last layer returning the most accurate labelling of the data.

\begin{figure}[!htb]
\center{\includegraphics[width=1\columnwidth]{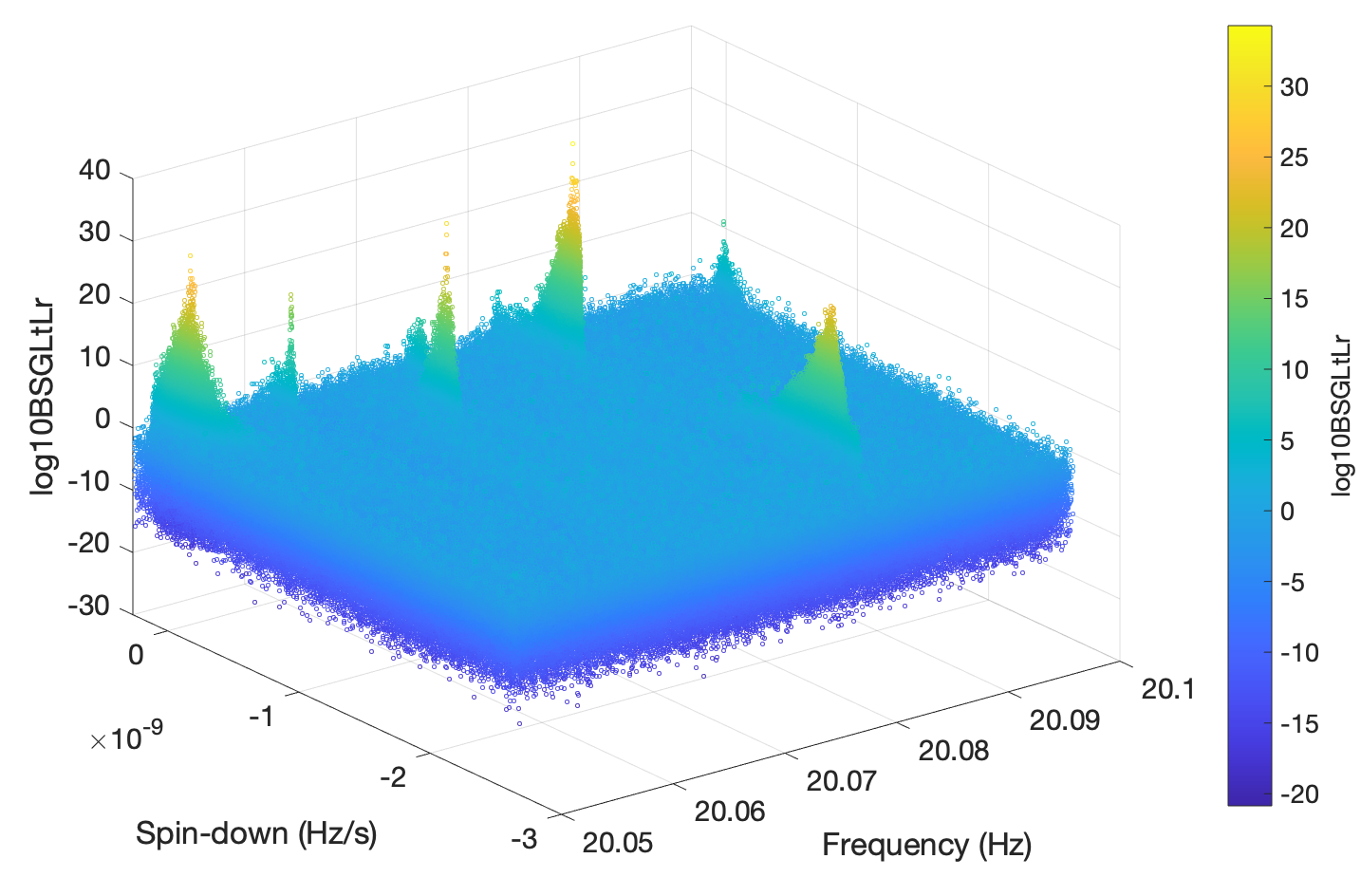}}
\caption{\label{cf1} Results in the 20.05-20.1 Hz band : the detection statistic as a function of $(f,\fdot)$. Several fake signals were added to the data before the search and appear in this result-set as parameter space regions with enhanced values of the detection statistic.}
\end{figure}

A simple network architecture includes an input layer, an output layer, and a middle layer which is called hidden layer, as schematically shown in Fig.~\ref{fig:NN}. More complex structures are possible with multiple hidden layers or internal level networks, merged in a larger network. A neural network is called deep when  it has more than two hidden layers \cite{sch14}. 

The training is performed starting from a random set of model weights. A few data sets are run through this initial network, the outputs are produced and compared with the corresponding ground truths. Based on this comparison the coefficients and biases are adjusted, and the network is updated. The process is repeated again with another batch of input data and corresponding ground truths. A new network with new model weights is produced. The network is trained through a series of successive trial-and-error adjustments, made by iterating this procedure a large number of times, to achieve an output that is closer and closer to the ground truth. The ``learning rate" describes by how much the model weights are modified at each training iteration.

The comparison between the output of a network and the corresponding ground truth is measured by the distance between the two, i.e. by the error made by the network. The training aims at minimizing such error by measuring the way in which the error changes when the model weights are changed and by  adjusting the latter accordingly. This is the optimisation problem at the heart of the functioning of neural networks. ``Loss" is also used to quantify error, and ``accuracy" is the complementary term to loss. 

Constructing a successful network is a combination of different factors: casting the labelling problem in an effective manner, preparing the input data to reflect most clearly the relationships that one wants to test, having a broad enough training data-set and determining the best network architecture (the structure of the layers, the optimisation algorithm, the activation function) and the best training parameters (e.g. the size of each training batch, the number of batches and the learning rate).

\subsection{Gravitational wave search results}
\label{cd}

We use as our test bed for the new clustering implementation, the results from the 
Einstein@Home search \cite{abb17}. This will make it possible to easily compare our results with the current clustering algorithm, used in \cite{abb17}. 

In an Einstein@Home search \cite{web1}  the computational workload is split in millions of separate tasks, called ``work units" (WUs) that are shared among the volunteer computers. Each task searches of order $10^9$ templates but only returns information about the most promising $10^4$ results. The information includes the value of the detection statistics and the associated template parameter values: $f,\fdot,\alpha,\delta$ which indicate frequency, frequency derivative, and the two sky-position coordinates right ascension and declination of the gravitational wave source. Typically each WU searches 50 mHz of signal frequencies, the entire $\fdot$ range and a subset of the sky. The results for the same 50 mHz frequency band from all regions of the sky are combined and constitute the result-set for each 50 mHz band. Fig.~\ref{cf1} shows the results in the 20.05-20.1 Hz band as an example.

\subsection{Preparation of the input data for the network}
\label{sec:NNdataPrep}
For each 50~mHz band of results-data we first produce a high-resolution image. The image has $\approx$ $60000\times22000$ ($f\times\fdot$) pixels, and the color associated with each pixel is determined by the value of the detection statistic. There are many empty pixels corresponding to templates whose detection statistic values were too low to be included in the set of results returned by the volunteer computer. There also exist  ($f$,$\fdot$) pixels where we find more than a candidate, corresponding to different sky positions. When this happens, we pick the one with the highest detection statistic value. Next we produce a lower resolution image by averaging over adjacent blocks of $40\times40$ pixels. This reduces the size of the image by  factor $>1500$ and reduces the number of empty pixels by a factor $\gtrsim 10$. Finally we divide this image into $256\times256$ pixels sub-images which can be handled by our 
high-end 32 GB GPU. This is the input to the network. Figure \ref{fa} shows a portion of a high resolution (original) image and the corresponding final sub-image for data that contain a fake signal.

\subsection{Network ground truth and training-set}
\label{subsec:groundTruthAndTrainingSet}

\begin{figure}[!htb]
\center{\includegraphics[width=0.8\columnwidth]{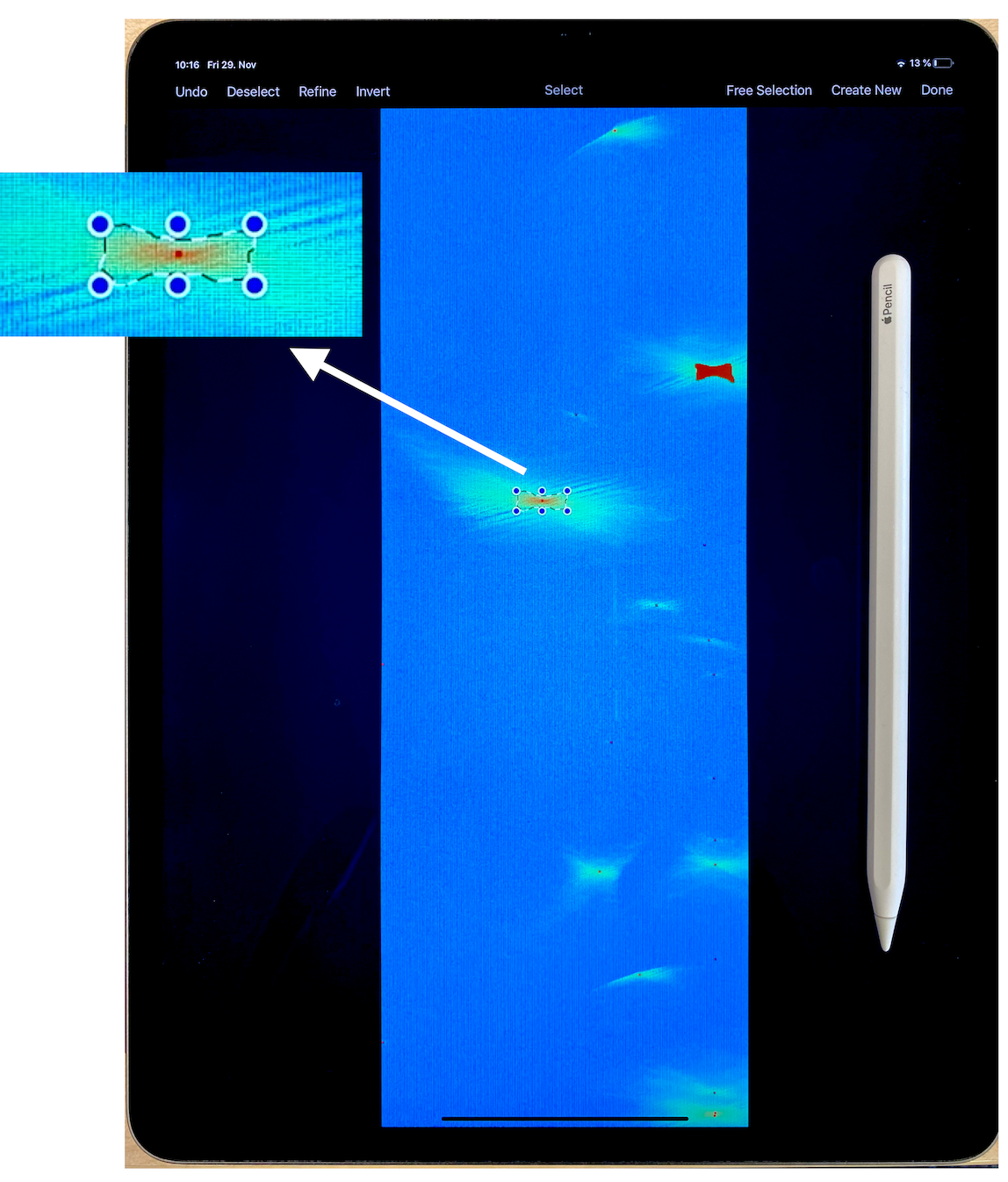}}
\caption{\label{ipad} A lower-resolution 50 mHz image on the tablet computer used to mark the ground-truth signal clusters. In this image 9 clusters are marked. The blue dotted-dashed-line (also the zoomed-in insert) shows the cluster region being identified and the brown area indicates a cluster which was previously identified and is already saved.}
\end{figure}

For each lower-resolution image the ground truth is a set of matrixes ${\cal{T}}_{ij}^\alpha$, with $(i,j)$ labelling the $(f,\fdot)$ pixels of the sub-image and $\alpha=1\cdots N_{\textrm{cl}}$ the total number of clusters found in that image. ${\cal{T}}_{ij}^\alpha=1$ if that pixel is part of a cluster, and zero otherwise. To generate the ground truth, we use a series of result-sets created by running the search on data that contains fake signals. For every 50 mHz band we create the lower-resolution images. In each we identify by eye the visible clusters and mark them using an image editing tool (Pixelmator) on a tablet computer equipped with a touchscreen and a stylus (Fig.~\ref{ipad}). Each region is saved with the editing tool and then converted to ${\cal{T}}_{ij}^\alpha$. It takes about one minute per signal-cluster to mark and save the associated region. 
 
\begin{figure}[!htb]
\center{\includegraphics[width=0.5\columnwidth]{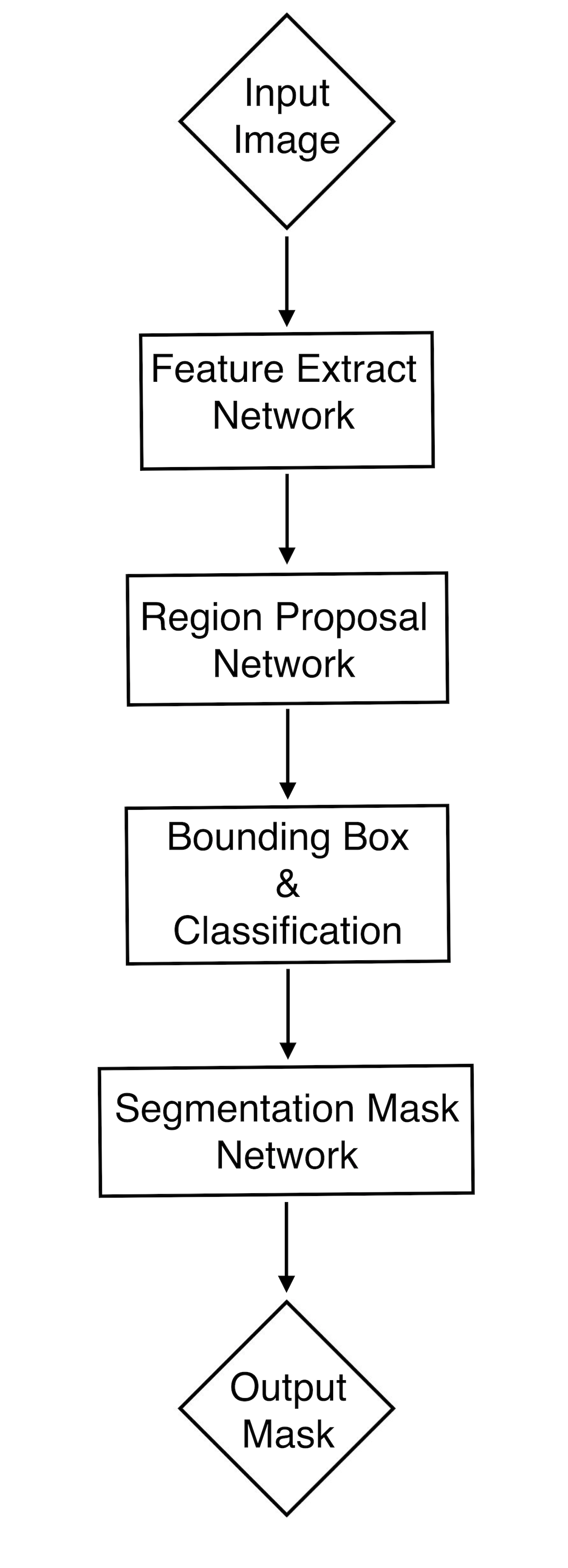}}
\caption{\label{n1} Schematic diagram showing the Mask RCNN architecture. }
\end{figure}

In order to train the network more thoroughly, we artificially augment the training-set by using the same sub-images again but having applied one or more of the following transformations on it: horizontal or vertical flip, rescaling all pixel values with the same random value (sampled uniformly from the interval 0.1-1.5 ) and dropping out a small fraction (20\%) of the pixels. The corresponding ground-truth masks are consistently transformed.

\subsection{Network Architecture}
\label{net}

Instance segmentation architectures have greatly developed in the past few years and are widely used in visual-recognition problems \cite{ron15,pin15,dai16}. Instance segmentation detects objects, i.e. classifies individual objects and localizes each with a bounding box, and also labels each pixel in a fixed  set of categories. For complex images containing different objects, R-CNN (Region-based Convolutional Neural Networks, \cite{girshick2013rich}) are becoming a crucial pre-requisite to instance segmentation, because they can successfully identify and discriminate {\it {different}} objects ({\it{semantic}} segmentation). Instance segmentation can then be applied on each identified category. 

We use a Mask R-CNN architecture \cite{he17}  and follow its implementation in \cite{git17} with the python deep learning library Keras. 
The Mask R-CNN network consists of two main stages. The first stage scans the image and finds the regions that most likely include the objects (feature extraction + region proposal); the second stage classifies the regions and generates definitive boundaries (bounding box and classification + segmentation mask). The output is a pixel mask that determines the boundaries and a score that each identified region is a cluster. We set a threshold on this score at 0.5 that determines when we accept the region as a cluster. So for our specific problem the classification is rather simple -- signal cluster or not so. The boundaries and classification are provided by the output ${\cal{T}}_{ij}^\alpha$, which is our output mask. Figure \ref{n1} shows this network framework. 

For the feature extraction step we make standard choices and use the Resnet101 \cite{he15}  and the FPN (Feature Pyramid Network) \cite{lin16} deep convolutional networks.  As  Optimizer, we use Stocastic Gradient Descent. 

\begin{figure}[!htb]
\center{\includegraphics[width=1\columnwidth]{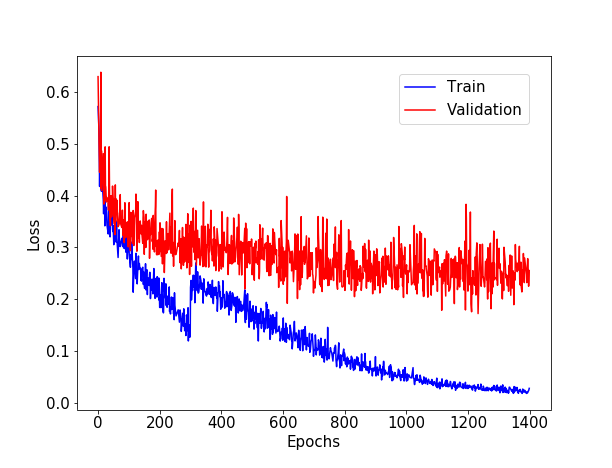}}
\caption{\label{acp} The network loss as a function of training epochs. The jump in the training happens at stage 3 when new untrained layers are added to the network.  }
\end{figure}

\subsection{Network training}

\begin{figure*}[hbt]
\vspace*{3mm}
\includegraphics[width=.85\columnwidth]{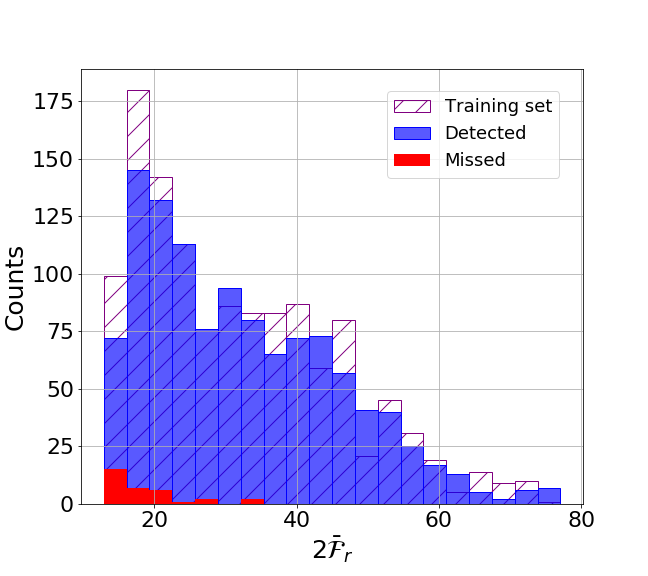}
\includegraphics[width=.85\columnwidth]{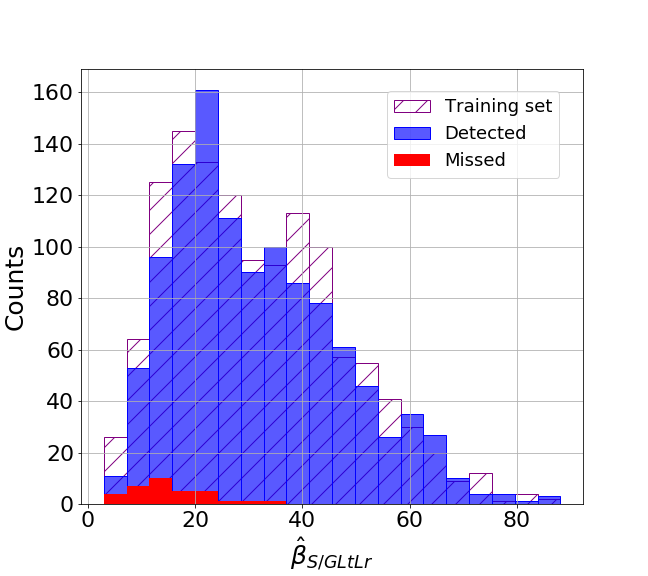}

\caption{\label{bs_test} Distribution of the detection statistics of the test-set (solid) and training-set (dashed) signals . For each signal we take as the representative candidate the one with the highest $\BSNtsc$$_r$ value. The right panel shows the distribution of the $\BSNtsc$$_r$ values for each candidate in the training- and test-set and the  left panel shows the distribution of their $2\avF_r$ values.  Some of the weaker test-set signals are not identified by the network; they are shown in red.}
\end{figure*}

The training is performed on a GPU Quadro GV100 with 32 GB memory. To enable the training on a larger batch size, our network framework allows a further reduction in the resolution of the input ground truth mask. In our case the input ground truth mask is internally always reduced to 28 $\times$ 28 pixels, represented by floats rather than zeros and ones, in order to hold more details. 

We use pre-trained model weights as a starting point for the network because preparing a large training-set ab initio is demanding. Although the pre-trained network has not been trained for our specific problem, using it significantly helps convergence of the training process while using a smaller training set. 

The training is performed in  3 steps: 1) only the first layer of the three last levels of the network (i.e. the three last networks of Fig.~\ref{n1}) are trained, for 100 epochs. 2) using the weights from the previous step, the first three levels are trained for 300 epochs 3) the complete network is trained with the weights from the second step for 1000  epochs. For the first two steps we use a learning rate=0.001. The learning rate in the third step is reduced to 0.0001. The best performance of the network is achieved with these hyper-parameters: batch size=15, weight decay=0.00001, learning momentum=0.9. 

Fig~\ref{acp} shows the evaluation of the network as a function of training epoch. The sudden increase in the loss shown in Fig.~\ref{acp} happens at the beginning of stage 3, when new layers with random weights are added to the training and so the network is in an unoptimized condition. The loss is determined by comparing the number of clusters identified by the network and the morphology of each cluster with the ground truth.

\section{Results}\label{res}

We train the network on 700 sub-images and test it on 670 sub-images. We refer to these data as the training-set and the test-set, respectively. The training-set only contains signal-clusters that can be identified by eye in the results images, i.e. loud signals with respect to the average level of the noise, as can be seen from Fig.~\ref{cf1}, and took about 28 hours to complete. There are 1245 clusters in the training-set and the distribution of the detection statistics for the loudest candidate of each cluster is shown in Fig.~\ref{bs_test}. For the interested reader, we show both the $\BSNtsc$$_r$ \cite{Keitel:2013wga, Keitel:2015ova} and $2\avF_r$ detection statistics because $\BSNtsc$$_r$ is the statistic that we use to rank the candidates, but the scaling of $2\avF_r$ with the signal amplitude is straightforward.

The trained network is applied to the test-set. The ground-truth for the test-set is established as done for the training set, i.e. with the procedure described in Section \ref{subsec:groundTruthAndTrainingSet}. From each cluster we take as representative the candidate with the highest value of the detection statistic. The distribution of candidate detection statistics 
is shown in Fig.~\ref{bs_test}, from which one can verify that the training-set and the test-set are comparable.

\begin{figure}[!h]
\vspace*{3mm}
\includegraphics[width=.7\columnwidth]{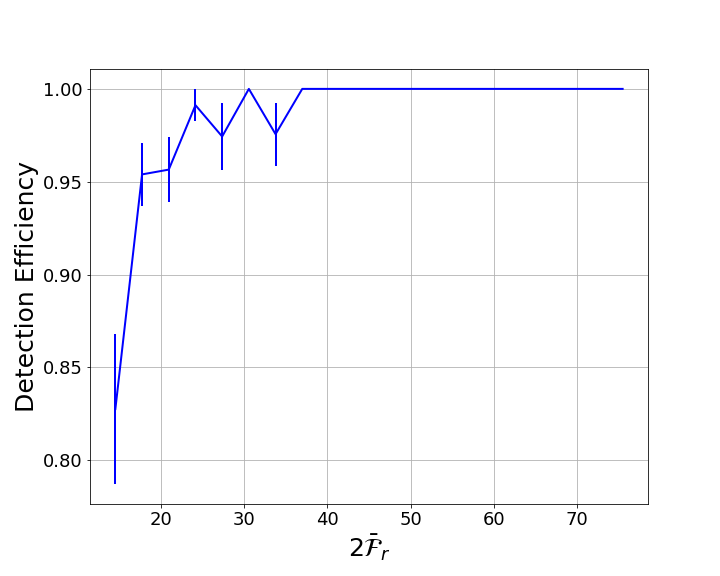}
\includegraphics[width=.7\columnwidth]{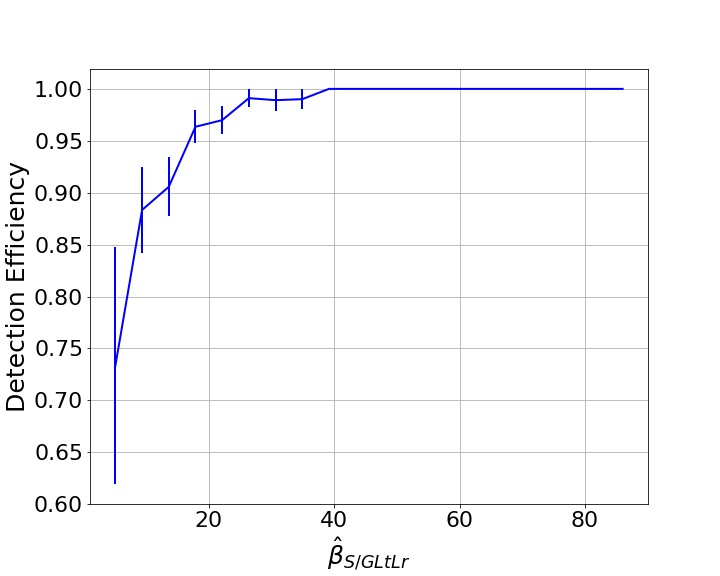}
\caption{\label{det} Detection efficiency of the network. The bars show the statistical error.  }
\end{figure}

The test-set comprises 1171 signal clusters. The network returns 1345 clusters of which 1137 are correctly identified signal clusters. This corresponds to a detection efficiency of  $\approx 97\%$. 
The network detection efficiency as a function of the candidate detection statistic is shown in Fig.~\ref{det}. All clusters with candidate $\BSNtsc$$_r$ above 35 or $2\avF_r$ above 40 are detected and the network has detection efficiency above 95\% at $\BSNtsc$$_r$  above 17 or $2\avF_r$ higher than 18. As a point of reference, for a population of signals at $\approx$ 90 Hz, with $h_0\approx 1.75\times 10^{-25}$, irrespective of whether they give rise to clusters which are like the ones that the network was trained for, the overall detection efficiency is $\sim 67\%$. $1.75\times 10^{-25}$ is lower than the upper limit at this frequency on O1 data from \cite{abb17}.

\begin{figure*}[hbt]
\includegraphics[width=0.6\columnwidth]{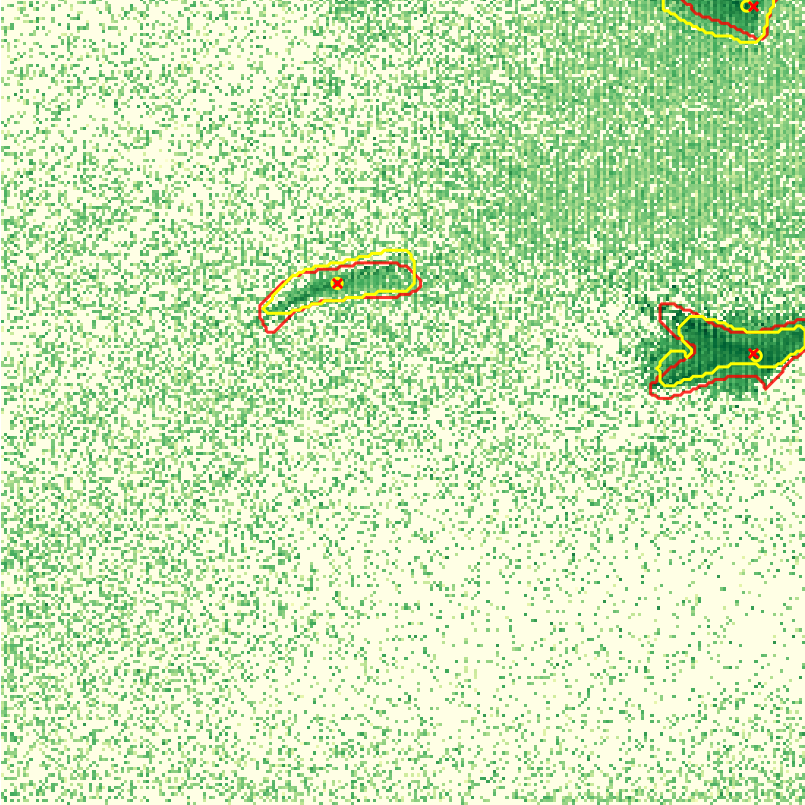}
\includegraphics[width=0.6\columnwidth]{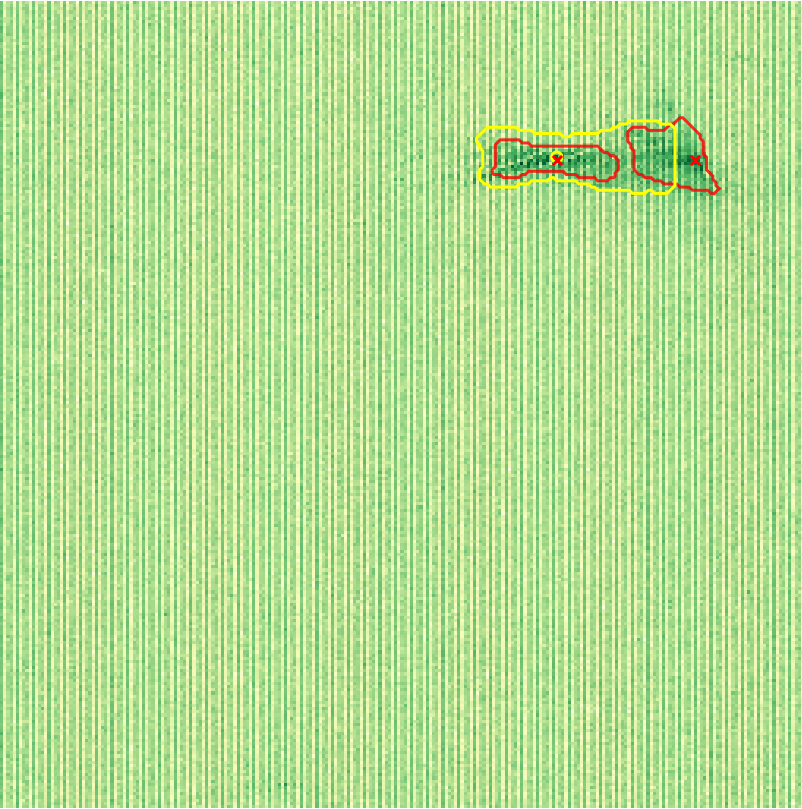}
\includegraphics[width=0.6\columnwidth]{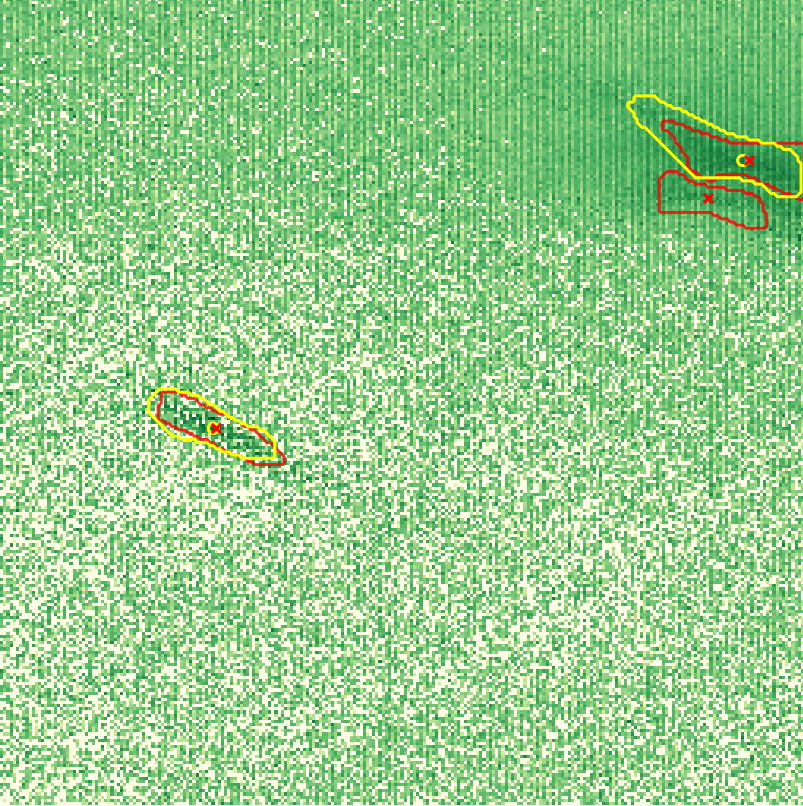}
\includegraphics[width=0.6\columnwidth]{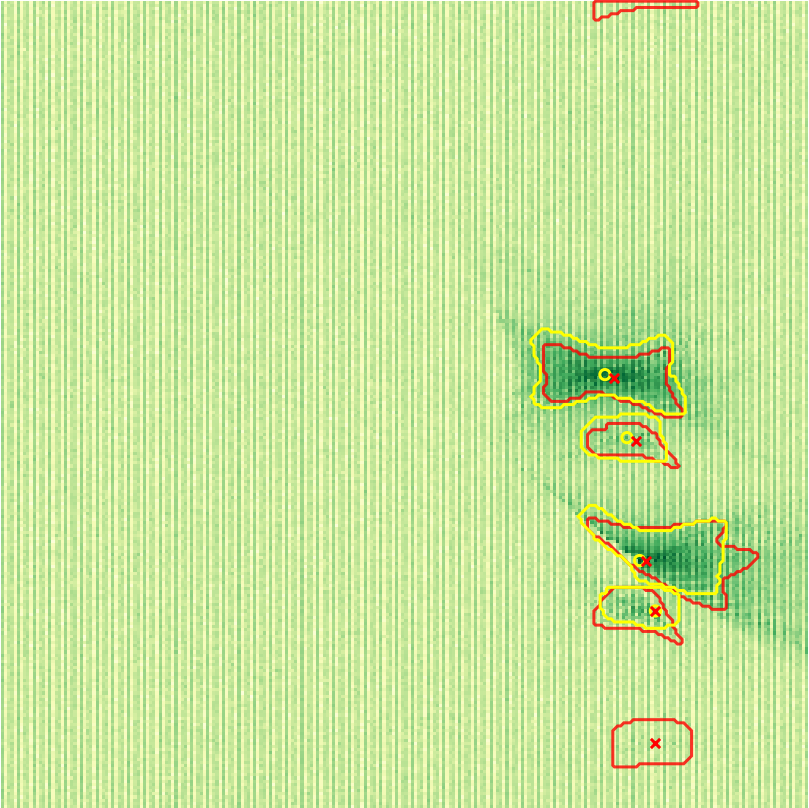}
\includegraphics[width=0.6\columnwidth]{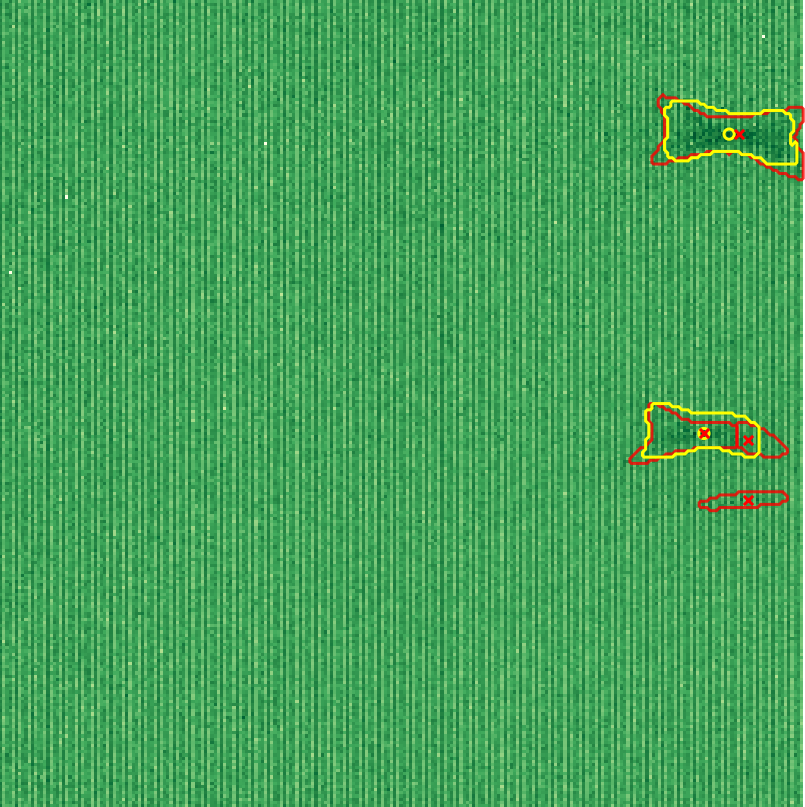}
\includegraphics[width=0.6\columnwidth]{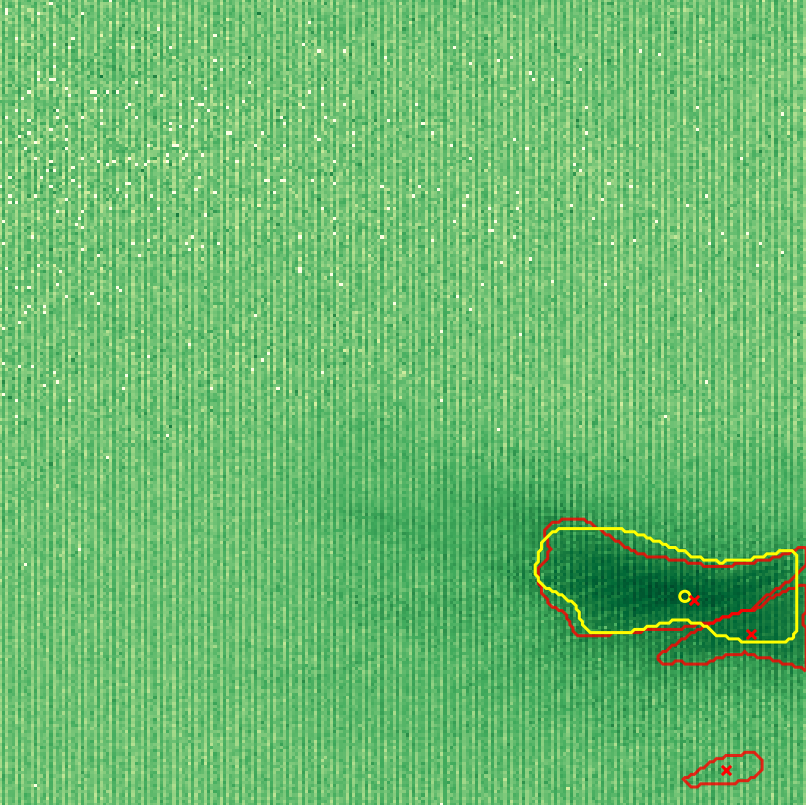}
\caption{\label{im_clus} Sample images from the test-set data : the color-coded detection statistic as function of signal frequency (vertical axis) and spin-down (horizontal axis). Images show the ground-truth clusters (red) and the clusters identified by the network on the same data (yellow). The red crosses mark the signal parameter values and the yellow circles mark the location in parameter space of the highest detection statistic candidate identified by the network. }
\end{figure*}

Fig.~\ref{im_clus} shows examples of clusters identified by the network on the test-set and the corresponding ground truth.
It is occasionally hard for the network to identify clusters which are close to larger clusters. In such situation either the two clusters are detected as one or are partially detected, or only the larger one is detected. The loss of detection efficiency for high-detection statistic value clusters stems only from this type of occurrence which is however not realistic for astrophysical signals because those are expected to be few and very sparse. At lower detection statistic values the small fraction of missed signal clusters is due to the signal features being too faint to be reliably identified.

\begin{figure}[!htb]
\center{\includegraphics[width=1.\columnwidth]{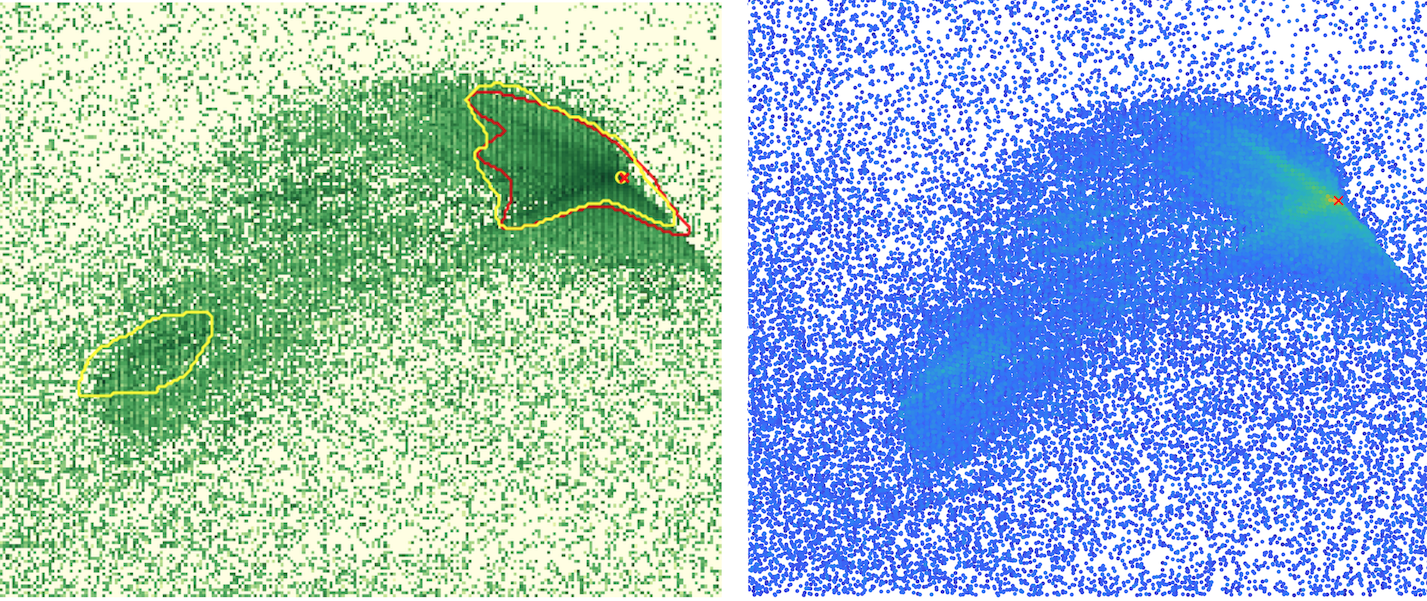}}
\caption{\label{fa} The frequency range between 23.15 Hz and 23.1585 Hz (vertical axis), the spin-down range between -10.8 and 2.64 $\times 10^{-10}$ Hz/s (horizontal axis) and color-coded the detection statistic values. The right panel shows a section from the original high resolution image; the left panel shows the same section from the lower resolution image. The red cross labels the fake signal parameter values and the red line the ground truth cluster. The yellow line shows the cluster identified by the network and the circle is its loudest candidate. A spurious cluster is visible in the left hand-side panel which is a by-product of the loud signal as explained in the text.}
\end{figure}

Most of the 174 ``false alarms'' are clusters that are very close to correctly-identified signal clusters. Fig.~\ref{fa} shows an example of this: a loud signal produces enhanced values of the detection statistic at parameter space points relatively far from the actual signal values. The network, quite rightly at this stage, identifies this as a signal cluster candidate and hence produces a ``spurious cluster". In an actual search we certainly do not expect many very loud signals, so this type of occurrence is not going to significantly contribute to the actual false alarm rate.
If we exclude these, the ``false alarms'' drop to 5\% of the total. 

\begin{figure}[hbt]
\vspace*{3mm}
\includegraphics[width=.7\columnwidth]{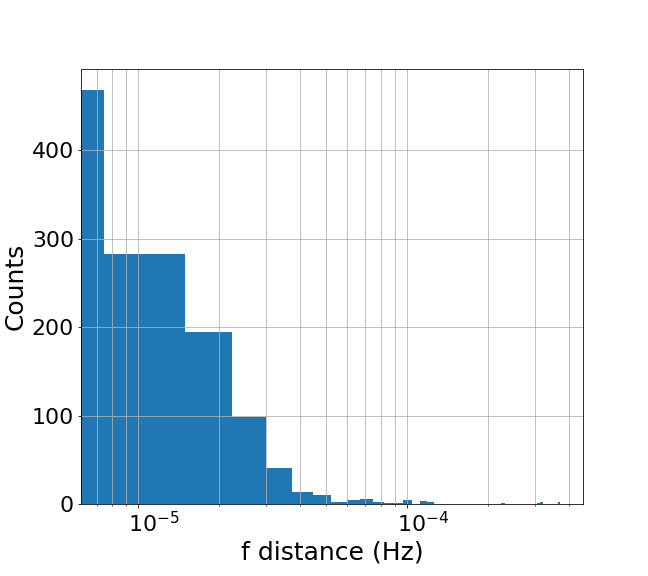}
\includegraphics[width=.7\columnwidth]{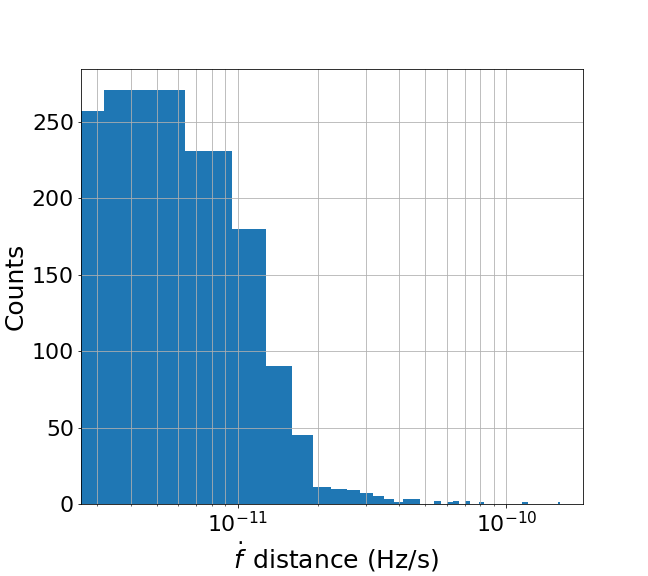}
\caption{\label{dist_dist} Distribution of the  distance between the signal parameters and the parameters of the most significant cluster candidate (the candidate with the highest $\BSNtsc$$_r$ value) recovered by the network. The top plot shows the distance in frequency; the bottom plot the distance in spin-down (right).}
\end{figure}

The loudest candidate associated with each signal cluster is considered and the distance of the template parameters from the actual signal parameters is determined, and shown in Fig.~\ref{dist_dist}. We find that the uncertainty in the signal parameters of this procedure is comparable to the uncertainty of the clustering algorithm \cite{abb17}: 90\% of the cluster-maxima are within $\approx10^{-4}$ Hz and $\approx10^{-11}$ Hz/s of the true signal parameter values. 

\section{Discussion and Summary}
\label{Section:summary}

We develop a Mask R-CNN to identify parameter space points that cluster with the morphology that we expect from a signal in the results of broad surveys for continuous gravitational waves. We train and test the network on the results from the \EatHs search \cite{abb17}. We concentrate on loud signals, that are visible by eye in the search results. 

Our network shows an excellent performance, detecting the large majority of the target-set signals even for detection statistic values ${{\BSNtsc}}$$_r =10$. The cluster-boundaries that the network defines are by and large consistent with the training-set and the distance of the cluster maximum from the nominal signal parameters consistent with the intrinsic uncertainties of the underlying search procedure. 

A typical Einstein@Home search produces of order $10^4$ 50-mHz result-sets, which typically are visually inspected by scientists to check for obvious signals. Albeit one learns a lot about the data and the results in this way,  $10^4$ is a large number, human attention degrades over time and so the reliability of this sort of procedure is hard to gauge in absolute terms. It would be preferable if this first cursory look at the results could be crossed-check with the results from a different type of system. Our neural network provides this and we will surely utilise it in the post-processing of the results from future \EatHs runs.

The question naturally arises of how the network performs on weaker signal. To answer this question we train the same network (i.e. without any optimization and without changing any of its hyperparameters) on much fainter signals. We find that the detection efficiency for these signals is much lower: for ${{\BSNtsc}}$$_r \sim 20$ the detection rate is $\sim$ 17\% and when $0 \leq {{\BSNtsc}}$$_r \leq 15$ it is $\approx$ 13\%. The reason is that the network works on images whose resolution is over 80 times lower than the resolution of the original result-set. This means that the network works on ``maxi-pixels'' each containing the average over nearly one hundred pixels of the original image. If the signal is weak, fewer pixels present elevated detection statistic values and the contrast that distinguishes the signal cluster is much degraded. With GPUs with a larger RAM it might be possible to use higher resolution data and improve the network performance for weaker signals.

Our clustering problem is intrinsically a multi-dimensional one. In the case of all-sky searches like \cite{abb17} it is 4D : frequency, frequency derivative and two sky coordinates. We have reduced it to a 2D problem by projecting all signal parameters in the frequency, frequency-derivative plane, as described in Section \ref{sec:NNdataPrep}. By doing so, we can cast our problem as an image processing problem, for which there is a vast literature and standard tools. A possible next step is to design a fully multi-dimensional cluster-recognition network.
   
We have also tested the ability of this network to identify large disturbances. Its performance turns out to be rather limited. This is not surprising because the network has been trained on signal-like morphologies, but interestingly it indicates that the network {\it{distinguishes}} loud signals from loud noise. We will leverage this and enhance the capabilities of the network to include the identification and classification of signals {\it and} disturbances. This will be extremely useful because it will allow to excise from the result-sets portions of disturbed parameter space while keeping the undisturbed ones for further analysis. Right now an automated procedure recognises 50 mHz bands that contain areas affected by disturbances and discards the entire 50 mHz as ``disturbed". The extension of this network to disturbances would allow to recuperate such regions.

\section{Acknowledgments}
The computing work for this project was carried out on the GPUs of the Atlas cluster of the Observational Relativity and Cosmology division of the MPI for Gravitational Physics, Hannover \cite{obsrelDiv}. We thank Bruce Allen for supporting this project by giving us access to those systems. We thank Reinhard Prix for his comments on the manuscript.

\bibliography{Bibliography}

\end{document}